\documentstyle[12pt]{article}

\catcode`\@=11
\long\def\@makefntext#1{ %\parindent 1em
\protect\noindent \hbox to 3.2pt {\hskip-.9pt
$^{{\ninerm\@thefnmark}}$\hfil}#1\hfill} %can be used

\def\thefootnote{\fnsymbol{footnote}}
 \def\@makefnmark{\hbox to 0pt{$^{\@thefnmark}$\hss}}  %original

\def\ps@myheadings{\let\@mkboth\@gobbletwo
\def\@oddhead{\hbox{} %\sl
\rightmark\hfil\ninerm\thepage}
\def\@oddfoot{}\def\@evenhead{\ninerm\thepage\hfil %\sl
\leftmark\hbox{}}\def\@evenfoot{}
\def\sectionmark##1{}\def\subsectionmark##1{}}

\begin{document}

\input epsf

%----------------------------PROCSLA.STY---------------------------------------
\newcommand{\symbolfootnote}{\renewcommand{\thefootnote}
	{\fnsymbol{footnote}}}
\renewcommand{\thefootnote}{\fnsymbol{footnote}}
\newcommand{\alphfootnote}
	{\setcounter{footnote}{0}
	 \renewcommand{\thefootnote}{\sevenrm\alph{footnote}}}

%------------------------------------------------------------------------------
%NEW DEFINED SECTION COMMANDS
\newcounter{sectionc}\newcounter{subsectionc}\newcounter{subsubsectionc}
\renewcommand{\section}[1] {\vspace{0.6cm}\addtocounter{sectionc}{1}
\setcounter{subsectionc}{0}\setcounter{subsubsectionc}{0}\noindent
	{\bf\thesectionc. #1}\par\vspace{0.4cm}}
\renewcommand{\subsection}[1] {\vspace{0.6cm}\addtocounter{subsectionc}{1}
	\setcounter{subsubsectionc}{0}\noindent
	{\it\thesectionc.\thesubsectionc. #1}\par\vspace{0.4cm}}
\renewcommand{\subsubsection}[1]
{\vspace{0.6cm}\addtocounter{subsubsectionc}{1}
	\noindent {\rm\thesectionc.\thesubsectionc.\thesubsubsectionc.
	#1}\par\vspace{0.4cm}}
\newcommand{\nonumsection}[1] {\vspace{0.6cm}\noindent{\bf #1}
	\par\vspace{0.4cm}}

%NEW MACRO TO HANDLE APPENDICES
\newcounter{appendixc}
\newcounter{subappendixc}[appendixc]
\newcounter{subsubappendixc}[subappendixc]
\renewcommand{\thesubappendixc}{\Alph{appendixc}.\arabic{subappendixc}}
\renewcommand{\thesubsubappendixc}
	{\Alph{appendixc}.\arabic{subappendixc}.\arabic{subsubappendixc}}

\renewcommand{\appendix}[1] {\vspace{0.6cm}
	\refstepcounter{appendixc}
	\setcounter{figure}{0}
	\setcounter{table}{0}
	\setcounter{equation}{0}
	\renewcommand{\thefigure}{\Alph{appendixc}.\arabic{figure}}
	\renewcommand{\thetable}{\Alph{appendixc}.\arabic{table}}
	\renewcommand{\theappendixc}{\Alph{appendixc}}
	\renewcommand{\theequation}{\Alph{appendixc}.\arabic{equation}}
%       \noindent{\bf Appendix \theappendixc. #1}\par\vspace{0.4cm}}
	\noindent{\bf Appendix \theappendixc #1}\par\vspace{0.4cm}}
\newcommand{\subappendix}[1] {\vspace{0.6cm}
	\refstepcounter{subappendixc}
	\noindent{\bf Appendix \thesubappendixc. #1}\par\vspace{0.4cm}}
\newcommand{\subsubappendix}[1] {\vspace{0.6cm}
	\refstepcounter{subsubappendixc}
	\noindent{\it Appendix \thesubsubappendixc. #1}
	\par\vspace{0.4cm}}

%------------------------------------------------------------------------------
%MARCO FOR ABSTRACT BLOCK
\def\abstracts#1{{
	\centering{\begin{minipage}{30pc}\tenrm\baselineskip=12pt\noindent
	\centerline{\tenrm ABSTRACT}\vspace{0.3cm}
	\parindent=0pt #1
	\end{minipage} }\par}}

%------------------------------------------------------------------------------
%NEW MACRO FOR BIBLIOGRAPHY
\newcommand{\bibit}{\it}
\newcommand{\bibbf}{\bf}
\renewenvironment{thebibliography}[1]
	{\begin{list}{\arabic{enumi}.}
	{\usecounter{enumi}\setlength{\parsep}{0pt}
%1.25cm IS STRICTLY FOR PROCSLA.TEX ONLY
\setlength{\leftmargin 1.25cm}{\rightmargin 0pt}
%0.52cm IS FOR NEW DATA FILES
%\setlength{\leftmargin 0.52cm}{\rightmargin 0pt}
	 \setlength{\itemsep}{0pt} \settowidth
	{\labelwidth}{#1.}\sloppy}}{\end{list}}

%------------------------------------------------------------------------------
%FOLLOWING THREE COMMANDS ARE FOR 'LIST' COMMAND.
\topsep=0in\parsep=0in\itemsep=0in
\parindent=1.5pc

%LIST ENVIRONMENTS
\newcounter{itemlistc}
\newcounter{romanlistc}
\newcounter{alphlistc}
\newcounter{arabiclistc}
\newenvironment{itemlist}
	{\setcounter{itemlistc}{0}
	 \begin{list}{$\bullet$}
	{\usecounter{itemlistc}
	 \setlength{\parsep}{0pt}
	 \setlength{\itemsep}{0pt}}}{\end{list}}

\newenvironment{romanlist}
	{\setcounter{romanlistc}{0}
	 \begin{list}{$($\roman{romanlistc}$)$}
	{\usecounter{romanlistc}
	 \setlength{\parsep}{0pt}
	 \setlength{\itemsep}{0pt}}}{\end{list}}

\newenvironment{alphlist}
	{\setcounter{alphlistc}{0}
	 \begin{list}{$($\alph{alphlistc}$)$}
	{\usecounter{alphlistc}
	 \setlength{\parsep}{0pt}
	 \setlength{\itemsep}{0pt}}}{\end{list}}

\newenvironment{arabiclist}
	{\setcounter{arabiclistc}{0}
	 \begin{list}{\arabic{arabiclistc}}
	{\usecounter{arabiclistc}
	 \setlength{\parsep}{0pt}
	 \setlength{\itemsep}{0pt}}}{\end{list}}

%------------------------------------------------------------------------------
%FIGURE CAPTION
\newcommand{\fcaption}[1]{
	\refstepcounter{figure}
	\setbox\@tempboxa = \hbox{\tenrm Fig.~\thefigure. #1}
	\ifdim \wd\@tempboxa > 6in
	   {\begin{center}
	\parbox{6in}{\tenrm\baselineskip=12pt Fig.~\thefigure. #1 }
	    \end{center}}
	\else
	     {\begin{center}
	     {\tenrm Fig.~\thefigure. #1}
	      \end{center}}
	\fi}

%TABLE CAPTION
\newcommand{\tcaption}[1]{
	\refstepcounter{table}
	\setbox\@tempboxa = \hbox{\tenrm Table~\thetable. #1}
	\ifdim \wd\@tempboxa > 6in
	   {\begin{center}
	\parbox{6in}{\tenrm\baselineskip=12pt Table~\thetable. #1 }
	    \end{center}}
	\else
	     {\begin{center}
	     {\tenrm Table~\thetable. #1}
	      \end{center}}
	\fi}

%------------------------------------------------------------------------------
%ACKNOWLEDGEMENT: this portion is from John Hershberger
\def\@citex[#1]#2{\if@filesw\immediate\write\@auxout
	{\string\citation{#2}}\fi
\def\@citea{}\@cite{\@for\@citeb:=#2\do
	{\@citea\def\@citea{,}\@ifundefined
	{b@\@citeb}{{\bf ?}\@warning
	{Citation `\@citeb' on page \thepage \space undefined}}
	{\csname b@\@citeb\endcsname}}}{#1}}

\newif\if@cghi
\def\cite{\@cghitrue\@ifnextchar [{\@tempswatrue
	\@citex}{\@tempswafalse\@citex[]}}
\def\citelow{\@cghifalse\@ifnextchar [{\@tempswatrue
	\@citex}{\@tempswafalse\@citex[]}}
\def\@cite#1#2{{$\null^{#1}$\if@tempswa\typeout
	{IJCGA warning: optional citation argument
	ignored: `#2'} \fi}}
\newcommand{\citeup}{\cite}

%------------------------------------GAC---------------------------------------
\newcommand{\be}{\begin{equation}}
\newcommand{\ee}{\end{equation}}
\newcommand{\bea}{\begin{eqnarray}}
\newcommand{\eea}{\end{eqnarray}}
\newcommand{\sumint}{\hbox{$\sum$}\!\!\!\!\!\!\!\int}
\newcommand{\cm}{M}
%------------------------------------GAC---------------------------------------
%------------------------------------------------------------------------------
%FOR FNSYMBOL FOOTNOTE AND ALPH{FOOTNOTE}
\def\fnm#1{$^{\mbox{\scriptsize #1}}$}
\def\fnt#1#2{\footnotetext{\kern-.3em
	{$^{\mbox{\sevenrm #1}}$}{#2}}}

%------------------------------------------------------------------------------
\font\twelvebf=cmbx10 scaled\magstep 1
\font\twelverm=cmr10 scaled\magstep 1
\font\twelveit=cmti10 scaled\magstep 1
\font\elevenbfit=cmbxti10 scaled\magstephalf
\font\elevenbf=cmbx10 scaled\magstephalf
\font\elevenrm=cmr10 scaled\magstephalf
\font\elevenit=cmti10 scaled\magstephalf
\font\bfit=cmbxti10
\font\tenbf=cmbx10
\font\tenrm=cmr10
\font\tenit=cmti10
\font\ninebf=cmbx9
\font\ninerm=cmr9
\font\nineit=cmti9
\font\eightbf=cmbx8
\font\eightrm=cmr8
\font\eightit=cmti8

%----------------------START OF DATA FILE------------------------------

\centerline{\tenbf FINITE TEMPERATURE EFFECTIVE POTENTIAL}
\baselineskip=22pt
\centerline{\tenbf IN THE HARTREE-FOCK APPROXIMATION}
\vspace{0.8cm}
\centerline{\tenrm Giovanni AMELINO-CAMELIA}
\baselineskip=13pt
\centerline{\tenit Center for Theoretical Physics, Laboratory
for Nuclear Science,}
\baselineskip=12pt
\centerline{\tenit and Department of Physics,
Massachusetts Institute of Technology}
\baselineskip=12pt
\centerline{\tenit Cambridge, Massachusetts 02139, USA}
\vspace{0.3cm}
\centerline{\tenrm and}
\vspace{0.3cm}
\centerline{\tenrm So-Young PI }
\baselineskip=13pt
\centerline{\tenit Department of Physics,
Boston University, 590 Commonwealth Ave.}
\baselineskip=12pt
\centerline{\tenit Boston, Massachusetts 02215, USA}
\vspace{0.9cm}
\abstracts{In order to investigate the nature
of the phase transition, we study the finite temperature
effective potential for the $\lambda \Phi^4$ theory in the
Hartree-Fock approximation, which sums up all the daisy
and superdaisy diagrams.
}

\vfil
%\vspace{0.8cm}
\twelverm   %modified by CLee 23/07/93
\baselineskip=14pt
\section{Introduction}
Temperature induced symmetry-changing phase-transitions in quantum field
theory$^{1-3}$ are important ingredients in modern cosmological scenarios.
The approximate critical temperature of a given phase transition
can be determined by calculating the one-loop finite temperature
effective potential\cite{doja}.
However, cosmological
scenarios often rely on the detailed nature
of the phase transition.
In particular, it has
been recently  suggested\cite{mat} that
the observed baryon asymmetry might have been
generated at the
electroweak phase transition,
if this transition is of first
order, and that the rate of baryon number violation
at the electroweak  phase transition depends
exponentially on the expectation value of the Higgs field just below
the phase transition.
Unfortunately, when the temperature $T$ is near or above
the critical temperature $T_c$ the one-loop approximation
does not give a reliable
estimate of the
finite temperature effective potential $V_T(\phi)$.
The fact that
when $T \ge T_c$
the one-loop term restores
the symmetries which are {\it spontaneously broken}
by the tree-level potential at $T$=$0$,
tells us that at
the phase transition the
perturbative approach based on the ordinary loop expansion\cite{doja}
breaks down.
For example, in the $\lambda \Phi^4$ scalar theory the
tree level potential is (in terms of renormalized quantities)
\be
V^{tree} = - {m^2_R \over 2} \phi^2 + {\lambda_R \over 4!} \phi^4
{}~,
\ee

\noindent
and the one-loop contribution can be approximated by\cite{doja}

\be
V^{one-loop} \sim {\lambda_R T^2 \over 48} \phi^2
{}~,
\ee

\noindent
and therefore at
high temperatures
(i.e. $T \! \ge \! \sqrt{24 m_R^2/\lambda_R} \! \simeq \! T_c$)
$V^{one-loop} \ge V^{tree}$ for all $\phi \le T$.
By using power counting
it has been argued\cite{wei,hsu}
that the dominant high-temperature contributions come from
the infinite classes of {\it daisy} and {\it superdaisy}
diagrams\cite{doja,wei,hsu}
and that they are non-negligible for all  $\phi < T$ at $T\sim T_c$.
Using again power counting one can show\cite{hsu,firef} that
the improved approximation of $V_T(\phi)$
obtained by adding
the daisy and superdaisy contributions to the
one-loop result should
be reliable (even when $T \sim T_c$) for
all $\phi > g T$, where $g$ is the  (largest)
coupling constant of the theory ($g$=$\sqrt{\lambda}$
in $\lambda \Phi^4$ theory), and up to order $g^3$.

The recent interest in temperature-induced phase transitions,
due to the {\it electroweak baryogenesis} idea,
has motivated numerous attempts$^{5-11}$ to evaluate the
leading (and subleading) high-temperature
contributions from daisy and superdaisy diagrams.
Some of the authors have
calculated an ``improved
one-loop" effective potential in which the tree-level propagators are
replaced by temperature dependent effective propagators,
which were obtained by summing the dominant high-temperature
contributions from  infinite-series of certain classes of self-energy
graphs in perturbation theory.
When one considers only the leading corrections to the effective
propagators  all results are in agreement with each other.
However, there have been various disagreements when the
subleading corrections
to the effective propagators,
which are important in determining the detailed
nature of the phase transition,
are included.

The difficulties that arise
in the improved one-loop calculations are
due to the fact that the substitution of improved propagators in the
one-loop effective potential is an ad-hoc approximation.
One needs a consistent loop expansion of the effective potential in terms
of the full propagator, as the
one given by the Cornwall-Jackiw-Tomboulis (CJT)
formalism of the effective action
and potential for composite operators\cite{corn}.
Using the CJT formalism it is easy to see\cite{gacpi} that
the daisy and superdaisy resummed effective potential
(the sum of the one-loop and the leading daisy-superdaisy contributions)
is given by
\be
 V_T(\phi) \simeq V^{res}_T(\phi,G_0)
{}~,\label{hfa}
\ee
\be
V^{res}_T(\phi,G) \equiv
V_{cl}(\phi)
+{1 \over 2} \, \sumint_k \, \ln G^{-1}(k)
+ {1 \over 2} \, \sumint_k \, [D^{-1}(\phi;k) G(k) -1]
+ V^*_2(\phi,G)
{}~,\label{hfb}
\ee
\be
\biggl[{\delta V^{res}_T(\phi,G) \over \delta G}\biggr]_{G=G_0}
= 0
{}~,\label{hfc}
\ee
where
\be
\sumint_p ~ \equiv \,
T \sum^{\infty}_{n=-\infty} \int {d^3p \over (2 \pi)^3}
{}~,
\label{imfeyn}
\ee
and $V^{*}_2$ is given by the leading
two-loop contributions to the effective potential
for composite operators.

We shall present
the daisy and superdaisy resummed finite
temperature effective potential $V_T^{res}$
for the $\lambda \Phi^4$ theory
in the imaginary time formalism.
We evaluate numerically the exact $V_T^{res}$,
and we investigate the analytic structure of $V_T^{res}$
using an high-temperature expansion.
Instead of dropping various  finite and
divergent terms, as has been done often in the recent literature,
renormalization is carried out explicitly.

\section{$\lambda \Phi^4$ Theory}
\vspace*{-0.7cm}
\subsection{Daisy and Superdaisy Resummmed Effective Potential}
\vspace*{-0.35cm}

The Euclidean Lagrange density for the single scalar field with
$\lambda \Phi^4$ interaction is given by
\be
L = {1 \over 2} (\partial_{\mu} \Phi) (\partial^{\mu} \Phi)
+{1 \over 2} m^2 \Phi^2
+{\lambda \over 4!}  \Phi^4 ~.
\label{lf}
\ee

\noindent
The tree-level propagator is
\be
D(\phi;k) = {1 \over k^2 + m^2 + {\lambda \over 2} \phi^2 } ~,
\label{dfif}
\ee

\noindent
and the vertices of the shifted ($\Phi \rightarrow \Phi + \phi$)
theory are given by
\be
L_{int}(\phi;\Phi) =
{\lambda \over 6} \phi \Phi^3 + {\lambda \over 4!} \Phi^4 ~.
\label{sif}
\ee

Following Eqs.(\ref{hfa})-(\ref{hfc}) one finds that the daisy and superdaisy
resummed effective potential for the $\lambda \Phi^4$ theory
is given by\cite{gacpi}
\bea
V^{res}_T(\phi,G)&=&V_{cl}(\phi)+{1 \over 2} \, \sumint_k \,
\ln G^{-1}(k)
+ {1 \over 2} \, \sumint_k \, [D^{-1}(\phi;k) G(k) -1] \nonumber\\
& & + {3 \over 4!} \lambda \sumint_k G(k) \sumint_p G(p)
{}~. \label{gexpy}
\eea

\noindent
The last term on the r.h.s. of Eq.(\ref{gexpy}) is the contribution
of the double-bubble diagram\cite{gacpi},
which is the leading
two-loop contribution to the effective potential
for composite operators in $\lambda \Phi^4$ theory.

By stationarizing $V^{res}_T$ with respect to $G$ we obtain the gap
equation:
\be
G^{-1}(k) = D^{-1}(k)
+ \, {\lambda \over 2} \, \sumint_p G(p) ~.
\label{gapgfb}
\ee

\noindent
It is straightforward to show by iteration that Eq.(\ref{gapgfb})
generates all daisy and superdaisy diagrams
that contribute to the full propagator in ordinary
perturbation theory. This is called Hartree-Fock approximation\cite{corn}.

It is convenient to
take the following {\it Ansatz} for $G(k)$
\be
G(k)= {1 \over k^2 + \cm^2 }
{}~.
\label{gansfi}
\ee

\noindent
In Eq.(\ref{gansfi}) we have made no assumption on the form of $G(k)$;
in fact,
at this stage the ``effective mass" $\cm$ is an unknown function
of the momentum $k$ to be determined using Eq.(\ref{gapgfb}).
Substituting Eqs.(\ref{dfif}) and (\ref{gansfi}) in Eq.(\ref{gapgfb}),
one obtains
\be
\cm^2 = m^2 - {\lambda \over 2} \phi^2 - {\lambda \over 2} P(M)
{}~,
\label{gapth}
\ee

\noindent
where
\be
P(M) \, \equiv \, \sumint_k  \, {1 \over k^2 + \cm^2 }
{}~,
\label{gxxth}
\ee

\noindent
which implies that in this approximation
$\cm$ is momentum independent.

In terms of the solution $\cm(\phi)$ of Eq.(\ref{gapth}), the
daisy and superdaisy resummed effective potential takes the form
\be
V^{res}_T(\phi,\cm(\phi)) =V^{0}+
V^{I}+V^{II}
{}~,
\label{vsuma}
\ee
\be
V^{0} = {1 \over 2} m^2 \phi^2
+{\lambda \over 4!}  \phi^4
{}~,
\label{vsumb}
\ee
\be
V^{I} = {1 \over 2} \, \sumint_k \, \ln [k^2+\cm^2(\phi)]
{}~,
\label{vsumc}
\ee
\be
V^{II} = - {\lambda \over 8}  P(\cm(\phi)) P(\cm(\phi))
{}~,
\label{vsumd}
\ee

\noindent
where $V^0$, $V^I$, and $V^{II}$
are the classical, one-loop,
and two-loop contributions respectively.

\subsection{Renormalizing the Effective Potential}

The expression of $V^{res}_T(\phi,\cm(\phi))$ in (\ref{vsuma}) contains
divergent integrals. Moreover, due to the fact that our
approximation is self-consistent,
reflecting the non-linearity of
the full theory, $\cm(\phi)$, the argument of $V^{res}_T$, is not
well-defined because of the infinities in $P(M)$.
We shall
first obtain a well-defined, finite expression for $\cm(\phi)$ by a
renormalization.

We define renormalized parameters $m_R$ and $\lambda_R$ as
\bea
\pm {m_R^2 \over \lambda_R} &=& {m^2 \over \lambda} +
{1 \over 2} I_1
{}~,
\label{renmla}\\
{1 \over \lambda_R} &=& {1 \over \lambda} +
{1 \over 2} I_2(\mu)
{}~,
\label{renmlb}
\eea

\noindent
where $m_R^2 > 0$, and $I_{1,2}$ are divergent integrals
\be
I_1 \equiv \int {d^3k \over (2 \pi)^3}~ {1 \over 2 |{\bf k}|} =
\lim_{\Lambda \rightarrow \infty} {\Lambda^2 \over 8 \pi^2}
{}~,
\label{renmlc}
\ee
\be
I_2(\mu) \equiv \int {d^3k \over (2 \pi)^3}~
[ {1 \over 2 |{\bf k}|} - {1 \over 2 \sqrt{|{\bf k}|^2 + \mu^2}} ] =
\lim_{\Lambda \rightarrow \infty} {1 \over 16 \pi^2}
\ln {\Lambda^2 \over \mu^2}
{}~.
\label{renmld}
\ee

\noindent
$\mu$ is the renormalization scale and $\Lambda$ is the ultraviolet
momentum cut-off.
In the following we shall choose the negative sign in Eq.(\ref{renmla}),
which allows spontaneous symmetry breaking.

When the sum on $n$ is carried out as in Ref.2, $P(M)$ takes
the form
\bea
P(M) &=& \int {d^3k \over (2 \pi)^3}~
{1 \over 2 \omega_k } + \int {d^3k \over (2 \pi)^3}~
{1 \over \omega_k (exp[\beta \omega_k]-1)}   \nonumber\\
&\equiv& P_f(\cm) + I_1 - \cm^2 I_2(\mu)  ~,
\label{gxxb}
\eea

\noindent
where $\omega_k \equiv [|{\bf k}|^2 + \cm^2]^{1/2}$ and
$P_f(\cm)$ is the finite part of $P(M)$, given by
\be
P_f(\cm) \equiv  {\cm^2 \over 16 \pi^2}
\ln {\cm^2 \over \mu^2} +
\int {d^3k \over (2 \pi)^3}~
{1 \over \omega_k (exp[\beta \omega_k]-1)}  ~.
\label{gom}
\ee

\noindent
In the limit $T=0$, the first term in $P_f(\cm)$ survives, but
the second term vanishes.

It is straightforward to see that $\cm$ is finite and cut-off
independent in terms of $m_R$ and $\lambda_R$
\be
\cm^2 =
- m_R^2 + {\lambda_R \over 2} \phi^2
+ {\lambda_R \over 2} P_f(\cm)
\equiv \tilde{m}^2(\phi)
+ {\lambda_R \over 2} P_f(\cm)
{}~,
\label{mgapfi}
\ee

\noindent
where we also defined, for later convenience,
the tree-level effective mass $\tilde{m}(\phi)$.

With this finite $\cm$, we are ready to discuss the
divergences in $V^{res}_T(\phi,\cm)$. First, carrying out the sum
on $n$ in $V^I$, we obtain the familiar one-loop finite
temperature formula\cite{doja}
\bea
\lefteqn{V^I(\cm) = {1 \over 2} \int {d^3k \over (2 \pi)^3}~
 \omega_k +
{1 \over \beta} \int {d^3k \over (2 \pi)^3}~
\ln (1-exp[\beta \omega_k])} ~~~   \nonumber\\
&=& \!\! {\cm^4 \over 64 \pi^2}
[\ln {\cm^2 \over \mu^2} - {1 \over 2}]
+ {1 \over \beta} \int {d^3k \over (2 \pi)^3}~
\ln (1-exp[\beta \omega_k])
- {\cm^4 \over 4}  I_2(\mu)
+{\cm^2 \over 2}  I_1
{}~. ~~~~~~
\label{viom}
\eea

\noindent
At $T=0$ the first term  of $V^I$ survives and provides the zero-temperature
one-loop contribution, and the second term vanishes. The last two
terms are the divergence in $V^I$.

Divergences in the two-loop contribution $V^{II}$ come from
the square of $P(M)$. Finiteness of $V^{res}_T$ can be shown by
first combining $V^{0}$ and $V^{II}$ using the unrenormalized form of
the gap equation. When the combined expression is written in terms of
renormalized parameters, the remaining divergent integrals are cancelled
by those of $V^{I}$ in (\ref{viom}). This is another indication that the
two-loop contribution must be included for a finite self-consistent
approximation. The resulting finite expression for
$V^{res}_T$ is
\be
V^{res}_T(\phi,\cm(\phi)) = (V^{0}_R+V^{II}_R)
+V^{I}_R ~,
\label{vrensuma}
\ee
\be
V^{0}_R+V^{II}_R = {\cm^4 \over 2 \lambda_R}
- {1 \over 2} \cm^2 P_f(\cm)
- {\lambda \over 12}  \phi^4
{}~,
\label{vrensumb}
\ee
\be
V^{I}_R = {\cm^4 \over 64 \pi^2}
[\ln {\cm^2 \over \mu^2} - {1 \over 2}] +
{1 \over \beta} \int {d^3k \over (2 \pi)^3}~
\ln (1-exp[\beta \omega_k]) ~.
\label{vrensumc}
\ee

\noindent
[A constant term $m^4/(2 \lambda)$ has been adjusted to obtain
Eqs.(\ref{vrensuma})-(\ref{vrensumc})
from Eq.(\ref{vsuma}).]
In order to compare the
finite temperature effective potential with and without the two-loop
contribution in our later discussion, we still have to extract $V^{II}_R$
from (\ref{vrensumb}). Observing that $V^0_R$ should be a function of
$\phi$ only, and
that in our approximation $V^{II}_R$ does not depend on $\phi$
explicitly [since the double-bubble graph
does not involve any vertices that depend on $\phi$]
one obtains, by using the renormalized gap equation,
\be
V_R^{0}+V_R^{II} = \biggl[ {\lambda_R \over 8}
\biggl( \phi^2- 2 {m_R^2 \over \lambda_R} \biggr) -
{\lambda \over 12} \phi^4 \biggr] -
{\lambda_R \over 8} P_f(\cm) P_f(\cm) ~.
\label{vovii}
\ee

\noindent
Clearly the last term in Eq.(\ref{vovii}) is the two-loop contribution.
The quantity in the brackets is the classical contribution after
renormalization is carried out. It is cut-off dependent
because of
the term $- \lambda \phi^4 /12$, which did not get renormalized due
to the structure of the gap equation.
But the renormalization
prescription (\ref{renmla})-(\ref{renmlb}) tells us that if
$\lambda_R$ is held fixed as $\Lambda \rightarrow \infty$, $\lambda$
approaches $0_-$. [A necessary condition in a non-trivial
($\lambda_R > 0$) renormalized $\lambda
\phi^4$ theory is $\lambda < 0$.] As shown in large N
studies\cite{momo}, such theory is intrinsically unstable.
On the other
hand, holding $\lambda > 0$ implies $\lambda_R \rightarrow 0$ as
$\Lambda \rightarrow \infty$. For $\lambda > 0$, a sensible
theory can be obtained for a fixed small $\lambda_R > 0$ as an
effective low energy theory, if $\Lambda$ is kept fixed at a large but
finite value.
Such theory requires
\be
{\lambda_R \over 32 \pi^2} \ln {\Lambda^2 \over \mu^2} < 1
{}~,
\label{condapi}
\ee

\noindent
in order to have $\lambda >0$, and all momenta, temperature and any
other physical mass scale must be much smaller than $\Lambda$.
We shall consider such an effective theory.

\begin{figure}[htb]
\epsfxsize=2.8in
\centerline{\epsffile{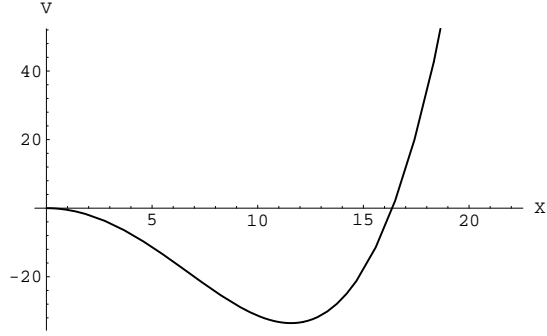}}
\caption{The daisy and superdaisy resummed effective potential
at $T$=0.
In figure $V(X) \! \equiv \! Re[V^{res}_T(X) \! - \! V^{res}_T(0)]/m_R^4$,
$X \! \equiv \! \phi / m_R$, $\lambda_R \! = \! 0.05$,
$\mu \! = \! m_R$, and $\ln (\Lambda^2/ m_R^2) = 16 \pi^2$.}
\label{fig3p7}
\end{figure}

As shown in Fig.\ref{fig3p7},
the zero-temperature phase structure of the effective
theory with finite $\Lambda$ is similar to that of perturbation theory:
there exists a minimum at a non-zero value of $\phi$.

%\begin{figure}[htb]
%\centerline{\psfig{figure=grbanfpi1.eps,height=2.5in}}
%\caption{The daisy and superdaisy resummed effective potential
%in $\lambda \Phi^4$ theory at $T$=0.
%$V^{res}_T$ illustrated in figure corresponds
%to $\lambda_R = 0.05$ and $\ln (\Lambda^2/ m_R^2) = 16 \pi^2$.
%$V^{res}_T$ becomes imaginary for small $\phi / m_R$.}
%\label{fig3p7}
%\end{figure}

\begin{figure}[htb]
\epsfxsize=2.8in
\centerline{\epsffile{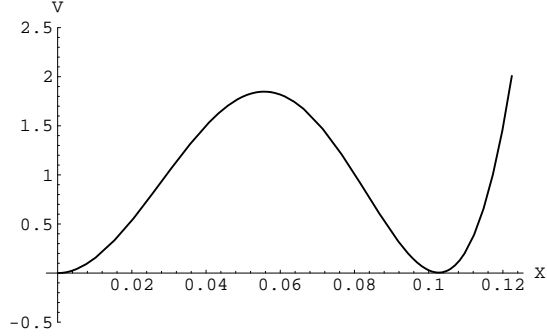}}
\caption{The daisy and superdaisy resummed
finite temperature effective potential.
In figure
$V(X) \! \equiv \! 10^{11} Re[V^{res}_T(X) \! - \! V^{res}_T(0)]/T^4$,
$X \! \equiv \! \phi / \lambda^{1/2}_R T$,
$T \! \simeq T_c \! \simeq \! 21.920 m_R$,
$\lambda_R = 0.05$,
$\mu = m_R$, and $\ln (\Lambda^2/ m_R^2) = 16 \pi^2$.}
\label{figp7th}
\end{figure}

The numerical result for the effective potential in Eq.(\ref{vrensuma})
(with finite $\Lambda$)
at the critical temperature is reported in Fig.\ref{figp7th}.
Notice that at the critical temperature the daisy and superdaisy
resummed effective potential has two degenerate minima.
However, the symmetry
breaking minimum $\phi=\phi_b$
is very close to the symmetric minimum $\phi =0$;
in fact, $\phi_b/T_c << \sqrt{\lambda_R}$.
As discussed in the introduction, the daisy and superdaisy
resummed effective potential is expected to give a reliable approximation
of the full effective potential only for $\phi > \sqrt{\lambda_R} T$, and
therefore,
since $\phi_b$ is located in the region
that is unreliably investigated by $V_T^{res}$,
the correct conclusion to be drawn from Fig.\ref{figp7th}
is that the phase transition of the $\lambda \Phi^4$ theory is
second order or very weakly first order.
Recent numerical investigations seem to indicate that
this phase transition is second order.

\subsection{High Temperature Approximation}

Our main interest is in the form of the effective
potential at high temperature (of the order of the critical temperature).
At high temperature one can derive an analytic approximation
of the daisy and superdaisy
resummed effective potential obtained in Eq.(\ref{vrensuma})
by assuming that $\cm^2/T^2 << 1$
in the integral expressions of $P_f(\cm)$ and $V_R^I(\cm)$.

As shown in Ref.2, for $\cm^2/T^2 << 1$
\be
P_f(\cm) \simeq T^2
\biggl[{1 \over 12 } - { 1 \over 4 \pi} {\cm  \over T } \biggr]
 ~.
\label{gomt}
\ee

\noindent
Then the high-temperature gap equation takes the form
\be
\cm^2 \simeq \tilde{m}^2(\phi)
+ {\lambda_R \over 24} T^2 -
{\lambda_R \over 8 \pi} \cm T
{}~.
\label{mgapfit}
\ee

\noindent
{}From the solution of this equation one finds that for a small coupling
$\lambda_R << 1$, the condition $\cm^2/T^2 << 1$ is consistent
with $\tilde{m}^2(\phi)/T^2 << 1$,
which is exactly the required condition for the high-temperature expansion
of the perturbative calculation in Ref.2.

Now we return to Eqs.(\ref{vrensuma})-(\ref{vovii}).
Using Eq.(\ref{mgapfit}) and the high-temperature expansion of
$V_R^I(\cm)$ derived in Ref.2,
we obtain the desired high-temperature analytic approximation
of the daisy and superdaisy
resummed effective potential:
\be
V^{res}_T(\phi,\cm(\phi)) = V^{0}_R+V^{I}_R+V^{II}_R
{}~,
\label{rewa}
\ee
\be
V^0 = {\lambda_R \over 8}
\biggl[\phi^2- 2 {m_R^2 \over \lambda_R}
\biggr]^2 - {\lambda \over 12}  \phi^4
{}~,
\label{rewb}
\ee
\be
V^{I} \simeq - {\pi^2 \over 90} T^4 +
{\cm^2 T^2 \over 24} -
{\cm^3 T \over 12 \pi}
{}~,
\label{rewc}
\ee
\be
{}~~~~V^{II} \simeq - {\lambda_R \over 8}  \biggl[{T^4 \over 144}  -
{\cm T^3 \over 24 \pi}
+ {\cm^2 T^2 \over 16 \pi^2} \biggr]
{}~,
\label{rewd}
\ee

\noindent
where $\cm$ is analytic solution of the quadratic equation
(\ref{mgapfit}).

\begin{figure}[htb]
\epsfxsize=2.8in
\centerline{\epsffile{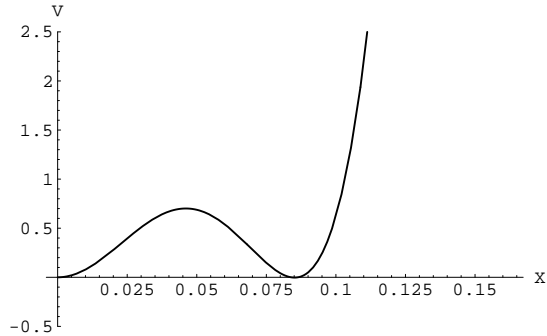}}
\caption{The high-temperature approximation
of the daisy and superdaisy resummed
effective potential.
In figure
$V(X) \! \equiv \! 10^{11} Re[V^{res}_T(X) \! - \! V^{res}_T(0)]/T^4$,
$X \! \equiv \! \phi / \lambda^{1/2}_R T$,
$T \simeq T_c \simeq 21.919 m_R$,
$\lambda_R = 0.05$,
$\mu = m_R$, and $\ln (\Lambda^2/ m_R^2) = 16 \pi^2$.}
\label{fig4p7}
\end{figure}

In Fig.\ref{fig4p7}, the  effective potential of
Eqs.(\ref{rewa})-(\ref{rewd}) is shown
at the critical temperature.
Note that the critical temperature that one obtains from the high-temperature
approximation of $V_T^{res}$ is
extremely close to the one obtained from the (numerical)
exact evaluation of $V_T^{res}$.
However, the high-temperature
approximation leads to a determination of the symmetry breaking
minimum $\phi_b$ which differs by approximately 20\%.

\section{Analysis of the Results}

In order to investigate
the structure of the high-temperature effective potential in
our approximation, we shall first consider the non-linear
aspects of the high-temperature gap equation
(\ref{mgapfit}),
which implies that $\cm(\phi)$ can be expanded for small $\lambda_R$ as
\be
\cm(\phi) = M_L(\phi)
\biggl\{ 1 - {\lambda_R T \over 16 \pi M_L(\phi) } +
O\biggl[ \biggl({ \lambda_R T \over 16 \pi M_L(\phi)} \biggr)^2 \biggr]
\biggr\}
{}~,
\label{maexpa}
\ee

\noindent
where
\be
M_L(\phi) \equiv \sqrt{\tilde{m}^2(\phi)
+ {\lambda_R \over 24} T^2}
\label{maexpb}
\ee

\noindent
solves the linearized high-temperature gap equation, i.e. (\ref{mgapfit})
without the last term on the r.h.s..

The one-loop contribution $V_R^I$ in Eq.(\ref{rewc})
can be written as
\be
V_R^{I}(\phi) \simeq - {\pi^2 \over 90} T^4 +
{M_L^2(\phi) T^2 \over 24} -
{\lambda_R  \over 192 \pi } \cm(\phi) T^3 -
{\cm^3(\phi) T \over 12 \pi}
{}~.
\label{vonma}
\ee

\noindent
The term linear in $\cm$, namely the third term on
the r.h.s. of Eq.(\ref{vonma}),
arises from the non-linearity of the
gap equation, i.e. from the last term on the r.h.s. of Eq.(\ref{mgapfit}).
If we were to use the linearized gap equation, without this term,
the first non-trivial correction to the
perturbative one-loop effective potential would be given by the term
cubic in $\cm$.
However, at high temperatures the leading non-linear correction is
of the same order as the term cubic in
$\cm$ in Eq.(\ref{vonma}); in fact, from Eq.(\ref{mgapfit}) we have
\be
\left({\lambda_R  \over 192 \pi } \cm(\phi) T^3 \right) \bigg/
\left({\cm^3(\phi) T \over 12 \pi}\right) \simeq {288 \over 192} \sim O(1)
\label{blublu}
\ee

\noindent
for $T >> \phi$.
When one includes the two-loop contribution given in
Eq.(\ref{rewd}) the $\cm T^3 $ term disappears and the high
temperature daisy and superdaisy resummed
effective potential in our approximation is
(neglecting some $\phi$-independent contributions)
\be
V_T^{res}(\phi) = V_R^{0}(\phi) +
\biggl({ T^2 \over 24} M_L^2(\phi) -
{T \over 12 \pi } M_L^3(\phi) \biggr) \biggl(1 + O(\lambda_R) \biggr)
+ O({\cm^4 \over 64 \pi^2 } \ln T) ~.
\label{gacpihf}
\ee

The above analysis of our consistent approximation shows that
improving the perturbative one-loop effective potential $V^I(\tilde{m})$
using the
non-linear gap equation clearly leads to an erroneous
result, and one must use a self-consistent method which relates the
effective potential and the gap equation.
However, we also
find that, due to the cancellation of the leading non-linear effect,  one
can obtain the leading daisy and superdaisy correction
by improving the one-loop
with a mass $M_L(\phi)$, which is the solution of the
linearized gap equation.
Such a procedure, which uses the effective mass squared shifted by
a $\phi$-independent amount proportional to $T^2$, was first suggested
by Weinberg\cite{wei} and later further studied by others\cite{car,fen}.

If one is also interested in the subleading daisy and superdaisy
correction,
a consistent result can only be obtained using the composite operator method
that we discussed.

\vglue 0.6cm
\leftline{\twelvebf Acknowledgements}
\vglue 0.4cm
This work is
supported in part by funds provided by the U.S. Department of Energy
(D.O.E.) under contract \#DE-FG02-91ER40676. G. A.-C. is supported by
the Istituto Nazionale di Fisica Nucleare,
Frascati, Italy.

\vglue 0.6cm
\leftline{\twelvebf References}
\vglue 0.4cm

\end{document}